\documentclass[a4paper,twocolumn,superscriptaddress,10pt,accepted=2018-12-31]{quantumarticle}
\pdfoutput=1
\usepackage[utf8]{inputenc}
\usepackage[english]{babel}
\usepackage[T1]{fontenc}

\usepackage{adjustbox}
\usepackage{amsmath}
\usepackage{epsfig}
\usepackage{floatrow}
\usepackage{hyperref}
\usepackage[numbers,sort&compress]{natbib}
\usepackage{physics}
\usepackage[binary-units]{siunitx}

\DeclareMathOperator{\qft}{QFT}
\DeclareMathOperator{\lcm}{lcm}

\begin{document}

\title{Optimising matrix product state simulations of Shor's algorithm}
\date{2019-01-09}

\author{Aidan Dang}
\email{aidan.dang@unimelb.edu.au}
\orcid{0000-0002-0047-5810}
\affiliation{Centre for Quantum Computation and Communication Technology, \\
    School of Physics, The University of Melbourne, Parkville, Victoria 3010,
    Australia}

\author{Charles D. Hill}
\email{cdhill@unimelb.edu.au}
\orcid{0000-0003-0185-8028}
\affiliation{Centre for Quantum Computation and Communication Technology, \\
    School of Physics, The University of Melbourne, Parkville, Victoria 3010,
    Australia}

\author{Lloyd C. L. Hollenberg}
\email{lloydch@unimelb.edu.au}
\orcid{0000-0001-7672-6965}
\affiliation{Centre for Quantum Computation and Communication Technology, \\
    School of Physics, The University of Melbourne, Parkville, Victoria 3010,
    Australia}

\maketitle

\begin{abstract}
We detail techniques to optimise high-level classical simulations of Shor's
    quantum factoring algorithm.
Chief among these is to examine the entangling properties of the circuit and to
    effectively map it across the one-dimensional structure of a matrix product
    state.
Compared to previous approaches whose space requirements depend on $r$, the
    solution to the underlying order-finding problem of Shor's algorithm, our
    approach depends on its factors.
We performed a matrix product state simulation of a 60-qubit instance of Shor's
    algorithm that would otherwise be infeasible to complete without an
    optimised entanglement mapping.
\end{abstract}

\section{Introduction}
With the potential for quantum computers to outperform the best classical
    computing resources available, achieving this quantum \textit{supremacy}
    promises to be a major milestone in computing.
However, the ability to demonstrate quantum supremacy
    \cite{Aaronson:2011:CCL:1993636.1993682, IQP, 2016arXiv160800263B} depends
    on several factors, including the computational task under consideration,
    as well as properties of the physical quantum computer (e.g. connectivity
    and error rate).
On the side of classical computation, the difficulty in defining a quantum
    supremacy point for a given problem stems from bounding the computational
    power of classical machines.
Therefore, we seek techniques to perform simulations of quantum algorithms and
    circuits more efficiently \cite{PhysRevLett.116.250501,
    2016arXiv160800263B, 2017arXiv171005867P}.
Generally, quantum algorithms possess some sort of structure that might be
    exploited to provide some advantage to classical simulation.
In this paper, we examine how the entanglement structure of Shor's algorithm
    for integer factorisation lends itself to a particular matrix product state
    representation that quantifiably reduces the computational requirements for
    classical simulation.
Additionally, we show how particular instances of Shor's algorithm become
    significantly easier to simulate when this structure is exploited.

Shor's algorithm \cite{doi:10.1137/S0097539795293172} may be used to factor a
    semiprime integer $N$ of $l$ binary digits in polynomial time with respect
    to $l$ on a quantum computer.
Presently, there is no known algorithm to perform this factorisation
    efficiently (i.e. in polynomial time) on a classical computer.
The presumed difficulty of factoring very large semiprime numbers is the
    foundation underpinning the security of the RSA public-key cryptosystem
    \cite{Rivest:1978:MOD:359340.359342} used to secure online communications
    \cite{RFC5246}.
This provides motivation to build quantum computers capable of performing large
    enough instances of Shor's algorithm and to develop public-key
    cryptosystems resistant to quantum computers \cite{Bernstein2009}.

Existing physical implementations of Shor's algorithm \cite{Vandersypen2001,
    PhysRevLett.99.250505, PhysRevLett.99.250504, Lucero2012} have been produced
    to factor small semiprimes no longer than five bits in length.
This is significantly less than even the 15-bit instances first simulated in
    \cite{Wang2017}, requiring 45 qubits.
Therefore, the simulated instances presented in this paper may be presented as
    medium-term goals for quantum hardware to benchmark against.

A typical state vector representation of an $n$-qubit system requires the
    storage of $2^n$ complex scalars, regardless of the state being stored.
While simulations of quantum circuits may be performed by operating on this
    collection of scalars, the exponential space complexity ultimately limits
    the size of the systems that can reasonably be simulated.
Instead, by using the matrix product state representation
    \cite{PhysRevLett.91.147902} of a quantum state, the space requirements
    scale according to the amount of entanglement present in the system.
As tensor networks \cite{ORUS2014117}, matrix product states were originally
    used for simulating one-dimensional quantum many-body systems
    \cite{SCHOLLWOCK201196, PhysRevLett.98.070201}, but have since been adapted
    for simulating quantum circuits \cite{PhysRevLett.91.147902, Wang2017,
    2014arXiv1406.0931W}.
As such, even states of many qubits may feasibly be stored using a matrix
    product state representation, provided its entanglement is sufficiently
    limited.
Other examples of tensor networks that may be used to simulate quantum circuits
    include PEPS \cite{peps}, MERA \cite{PhysRevLett.99.220405} and tree tensor
    networks \cite{PhysRevA.96.062322}.

By examining the entanglement introduced by a high-level circuit of Shor's
    algorithm, we were able to improve space requirements over previous
    simulations \cite{Wang2017} by sensibly mapping this entanglement across
    the one-dimensional structure of a matrix product state.
Furthermore, in treating the task of simulating Shor's algorithm as a sampling
    problem, we could take advantage of single-qubit measurement to collapse
    entanglement and hence reduce our space usage.
The combined result of our optimisations allowed us to simulate a nontrivial
    instance of Shor's algorithm involving 60 qubits on a supercomputer using
    an approximate total of $\SI{14}{\tera\byte}$ of RAM.
In comparison to the $2^{60} \times \SI{128}{\bit} \simeq
    \SI{1.8e7}{\tera\byte}$ to store a state vector for 60 qubits in double
    precision, this represents a significant reduction in the required memory.

To outline this paper, we begin with a review of Shor's algorithm in
    Sec.~\ref{sec:shors_algorithm} and a review of the matrix product state
    formalism in Sec.~\ref{sec:mps_review}.
We then examine the entanglement introduced at particular stages in Shor's
    algorithm in Sec.~\ref{sec:shors_entanglement} and then detail in
    Sec.~\ref{sec:mps_impl} how the subsystems of a matrix product state may be
    partitioned to take advantage of this entanglement evolution.
Finally, we provide benchmarks of our implementation in
    Sec.~\ref{sec:benchmarks}, including our simulation of a 60-qubit instance.

\section{Review of Shor's Algorithm} \label{sec:shors_algorithm}
\begin{figure*}[htb]
    \epsfig{file=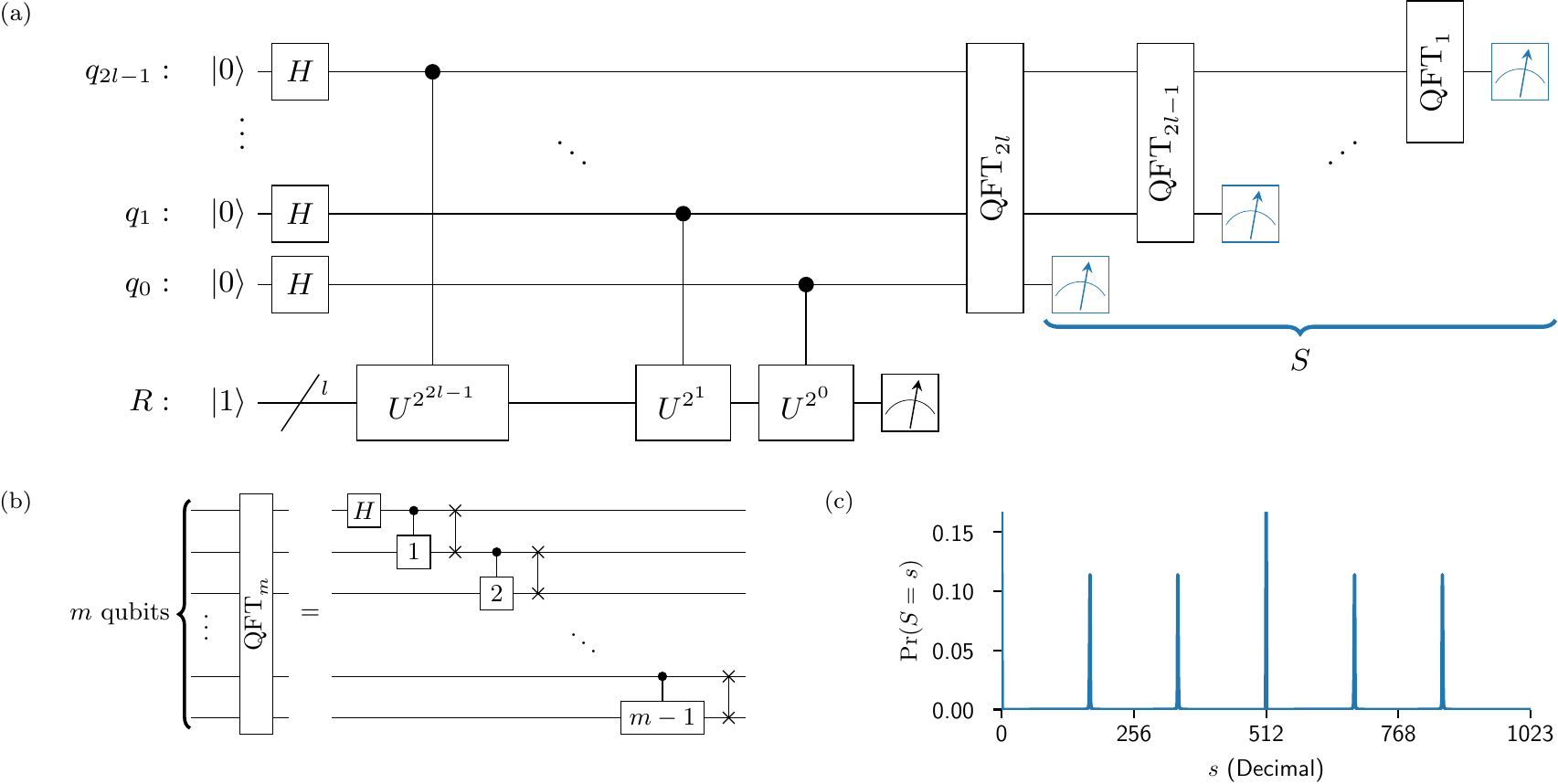,width=\textwidth,clip=}
    \caption{
        Schematic for the order-finding quantum subroutine in Shor's algorithm.
        (a) High-level circuit diagram for factoring a semiprime $N$ of $l$
            binary digits.
        For randomly chosen $a$ such that $1 < a < N$, $U$ performs $U \ket{x}
            \to \ket{ax \bmod N}$.
        $S$ is the result of final measurement of the upper register.
        (b) A single block in the linear nearest-neighbour quantum Fourier
            transform.
        A phase gate labelled by $x$ performs $\ket{0} \to \ket{0}$ and
            $\ket{1} \to \exp(-i \pi / 2^x) \ket{1}$.
        (c) Example probability distribution for $S$, where $l = 5$, $N = 21$,
            and $a = 2$.
    }
    \label{fig:shor}
\end{figure*}

We consider the circuit shown in Fig.~\ref{fig:shor} for this review of Shor's
    algorithm and for the general structure of our following simulations.
Given an odd, squarefree semiprime $N = p q$ represented by at least $l$ binary
    digits, a high-level circuit for Shor's algorithm as detailed in
    \cite{Nielsen:2011:QCQ:1972505} consists of an `upper' register of $2 l$
    qubits initialised to $\ket{0}$ and a `lower' register $R$ of $l$ qubits
    initialised to the state $\ket{1}$:
    \begin{align*}
        \ket{\psi_\text{init}} \equiv \ket{0} \otimes \ket{1}.
    \end{align*}
The Hadamard gate is then applied to each qubit of the upper register, creating
    the superposition
    \begin{align*}
        \ket{\psi_\text{superpos}} \equiv{}& (H^{\otimes 2 l}) \otimes
            (I^{\otimes l}) \ket{\psi_\text{init}} \\*
        &= \sum_{i = 0}^{2^{2 l} - 1} \ket{i} \otimes \ket{1},
    \end{align*}
    where we shall ignore normalisation constants for clarity.
Randomly choosing integer $a \neq 1$ from $\mathbf{Z}_N^*$, the set of integers
    modulo and coprime to $N$, exponents of the unitary operator
    \begin{align} \label{eq:modular_unitary}
        U\ket{x} \equiv \ket{a x \bmod N}
    \end{align}
    are applied to the lower register, controlled by qubits in the top:
    \begin{align} \label{eq:modular_exponentiation}
        \ket{\psi_\text{modexp}} \equiv{}& \sum_{i = 0}^{2^{2 l} - 1} \ket{i}
            \otimes U^i \ket{1} \nonumber \\*
        &{}= \sum_{i = 0}^{2^{2 l} - 1} \ket{i} \otimes \ket{a^i \bmod N}
            \nonumber \\*
        &{}= \sum_{\substack{j = 0 \\ k r + j < 2^{2 l}}}^{r - 1}
            \sum_{\substack{k = 0 \\ k < 2^{2 l} / r}} \ket{k r + j} \otimes
            \ket{a^j \bmod N},
    \end{align}
    where $r$ is the period of $U$.
To determine this period $r$, the lower register is measured, forcing a choice
    of the index $j$ in Eq.~\eqref{eq:modular_exponentiation}.
This collapses the entanglement between the two registers, and we can now
    separate them.
The upper register is now in state
    \begin{align} \label{eq:upper_register}
        \ket{\psi_\text{upper}} \equiv
            \sum_{\substack{k = 0 \\ k < 2^{2 l} / r}} \ket{k r + j}
    \end{align}
    for the $0 \leq j < r$ corresponding to the value $(a^j \bmod N)$ measured
    from the lower register in Eq.~\eqref{eq:modular_exponentiation}.

The quantum Fourier transform (QFT) is then applied to this upper register:
    \begin{align*}
        \qft \ket{\psi_\text{upper}} &= \sum_{s = 0}^{2^{2 l} - 1}
            \sum_{\substack{k = 0 \\ k < 2^{2 l} / r}}
            e^{2 \pi i (k r + j) s / 2^{2 l}} \ket{s} \\*
        &= \sum_{s = 0}^{2^{2 l} - 1} e^{2 \pi i j s / 2^{2 l}}
            \sum_{\substack{k = 0 \\ k < 2^{2 l} / r}}
            e^{2 \pi i k r s / 2^{2 l}} \ket{s}.
    \end{align*}
The upper register is then measured to produce a value $s$ with probability
    \begin{align} \label{eq:dist}
        \Pr(S = s) \propto \abs{\sum_{\substack{k = 0 \\ k < 2^{2 l} / r}}
            e^{2 \pi i k r s / 2^{2 l}}}^2
    \end{align}
    for random variable $S$.
This implies that values of $s$ such that $r s / 2^{2 l}$ is close to an integer
    are more likely to be measured, resulting in the peaks shown in the example
    probability distribution in Fig.~\ref{fig:shor}(c).
The quantum processing component of Shor's algorithm is now complete, and the
    result of $S$ may be classically processed using the continued fractions
    and Euclidean algorithms to successfully factor $N$ with high probability
    \cite{Nielsen:2011:QCQ:1972505}.

\section{Review of Matrix Product States} \label{sec:mps_review}
To briefly review the matrix product state (MPS) \cite{PhysRevLett.91.147902}
    formalism for representing quantum states, we begin with some general
    composite state of $n$ subsystems given by
    \begin{align} \label{eq:state_vector}
        \ket{\psi} = \sum_{i_1 = 0}^{d_1 - 1} \sum_{i_2 = 0}^{d_2 - 1} \cdots
            \sum_{i_n = 0}^{d_n - 1} \psi_{i_1 i_2 \ldots i_n} \ket{i_1} \otimes
            \ket{i_2} \otimes \cdots \otimes \ket{i_n}.
    \end{align}
A subsystem $m$ with dimensionality $d_m$ is referred to as a $d_m$-level
    \textit{qudit}.
In the case where $d_m = 2$, $m$ is a qubit.
Conventionally, the \textit{state-vector} approach for computationally storing
    this state involves recording each of the complex coefficients $\psi_{i_1
    i_2 \ldots i_n}$ in a single, perhaps multidimensional, array.
The space complexity of this particular representation is fixed at $O(d)$
    regardless of the specific state $\ket{\psi}$, where $d = \prod_{m = 1}^n
    d_m$.

At its core, the MPS representation of a state involves storing matrices for
    each qudit, whose product equals each of the coefficients in
    Eq.~\eqref{eq:state_vector}:
    \begin{align} \label{eq:basic_mps}
        \psi_{i_1 i_2 \ldots i_n} = \Gamma_{\alpha_1}^{\qty[1] i_1}
            \Gamma_{\alpha_1 \alpha_2}^{\qty[2] i_2} \cdots
            \Gamma_{\alpha_{n - 1}}^{\qty[n] i_n},
    \end{align}
    where we have used the usual summation notation.
Notably, qudits whose matrices are adjacent in Eq.~\eqref{eq:basic_mps} may be
    contracted and redefined into a higher-dimensional qudit:
    \begin{align} \label{eq:contraction}
        \Gamma_{\alpha_{m - 1} \alpha_{m + 1}}^{\qty[m, m + 1]
            \qty{i_m, i_{m + 1}}}
            \equiv \Gamma_{\alpha_{m - 1} \alpha_m}^{\qty[m] i_m}
            \Gamma_{\alpha_m \alpha_{m + 1}}^{\qty[m + 1] i_{m + 1}}.
    \end{align}
By sequentially contracting adjacent qudits, the state vector representation
    may be obtained from an MPS representation of a state.
Additionally, the reverse operation of Eq.~\eqref{eq:contraction} may be
    performed to decompose the matrices of two combined qudits.
This involves rearranging the elements of $\Gamma_{\alpha_{m - 1} \alpha_{m +
    1}}^{\qty[m, m + 1] \qty{i_m, i_{m + 1}}}$ in Eq.~\eqref{eq:contraction}
    and applying a matrix decomposition such as the trivial decomposition
    \begin{align} \label{eq:trivial_decomposition}
        M_{ab} = \begin{cases}
            M_{a \alpha_m} I_{\alpha_m b}, &\quad \dim \qty{a} \geq \dim \qty{b}
                \\*
            I_{a \alpha_m} M_{\alpha_m b}, &\quad \dim \qty{a} < \dim \qty{b}
        \end{cases},
    \end{align}
    where $I$ is the appropriately sized identity matrix, or the rank-revealing
    singular value decomposition (SVD)
    \begin{align} \label{eq:svd}
        M_{ab} = U_{a \alpha_m} S_{\alpha_m} V_{\alpha_m b},
    \end{align}
    where $U$ and $V$ are unitary matrices and the singular values, the elements
    of $S$, are nonnegative reals.
The intermediate index $\alpha_m$ is referred to as the bond index of its
    respective bipartition.
By sequentially decomposing qudits into constituent subsystems, an MPS
    representation of a state represented in the state vector form may be
    obtained.

Though the trivial decomposition, Eq.~\eqref{eq:trivial_decomposition}, is
    simple to perform, an MPS produced solely from such decompositions on a
    state vector representation offers no computational advantage in terms of
    time or space usage.
Instead, if the SVD, Eq.~\eqref{eq:svd}, is used as the decomposition, the MPS
    representation might look like
    \begin{align} \label{eq:mps}
        \psi_{i_1 i_2 \ldots i_n} = \Gamma_{\alpha_1}^{\qty[1] i_1}
            \lambda_{\alpha_1}^{\qty[1]}
            \Gamma_{\alpha_1 \alpha_2}^{\qty[2] i_2}
            \lambda_{\alpha_2}^{\qty[2]} \cdots
            \lambda_{\alpha_{n - 1}}^{\qty[n - 1]}
            \Gamma_{\alpha_{n - 1}}^{\qty[n] i_n},
    \end{align}
    where each $\lambda^\qty[m]$ is the vector of singular values obtained from
    Eq.~\eqref{eq:svd}.
If care is taken during the treatment of these singular values
    \cite{PhysRevLett.91.147902}, a canonical MPS form can be defined such that
    these $\lambda_{\alpha_m}^\qty[m]$ are equal to the Schmidt coefficients
    \cite{Nielsen:2011:QCQ:1972505} across bipartition $\qty[1, \ldots, m] :
    \qty[m + 1, \ldots, n]$.
Furthermore, if $\chi^\qty[m]$ is the Schmidt rank across this bipartition,
    $\dim \qty{\alpha_m}$ can be set to $\chi^\qty[m]$ by only storing values
    corresponding to nonzero Schmidt coefficients.
Consequently, with the Schmidt ranks as a measure of the entanglement within a
    state, `less entangled' states in a canonical MPS representation have lower
    space requirements to store.

To simulate measurements on qudit $m$ of a composite system, the reduced
    density matrix of $m$ may be obtained from the general MPS expression,
    Eq.~\eqref{eq:basic_mps}, by tracing over all other qudit subsystems:
    \begin{multline} \label{eq:density_contract}
        \rho_{i_m i'_m}^\qty[m] = \Gamma_{\alpha_1}^{\qty[1] i_1} \cdots
            \Gamma_{\alpha_{m - 1} \alpha_m}^{\qty[m] i_m} \cdots
            \Gamma_{\alpha_{n - 1}}^{\qty[n] i_n} \\*
        \times \overline{\Gamma_{\beta_1}^{\qty[1] i_1}} \cdots
            \overline{\Gamma_{\beta_{m - 1} \beta_m}^{\qty[m] i'_m}} \cdots
            \overline{\Gamma_{\beta_{n - 1}}^{\qty[n] i_n}},
    \end{multline}
    where the bars denote complex conjugation.
From this, the measurement probabilities of the $d_m$ possible values of $m$
    may be read from the diagonal elements of $\rho^\qty[m]$.
For a canonical MPS in the form of Eq.~\eqref{eq:mps}, $\rho^\qty[m]$ may
    instead be obtained locally:
    \begin{align} \label{eq:density_local}
        \rho_{i_m i'_m}^\qty[m] = \abs{\lambda_{\alpha_{m - 1}}^\qty[m - 1]}^2
            \Gamma_{\alpha_{m - 1} \alpha_m}^{\qty[m] i_m}
            \overline{\Gamma_{\alpha_{m - 1} \alpha_m}^{\qty[m] i'_m}}
            \abs{\lambda_{\alpha_m}^\qty[m]}^2,
    \end{align}
    due to unitarity of the resulting matrices of an SVD.
Even if an MPS is not in canonical form, the reduced density matrix
    $\rho^\qty[m]$ of qudit $m$ may still be obtained from
    Eq.~\eqref{eq:density_local} if pairwise contractions and SVDs are
    performed from the ends of the MPS in toward $m$.
We refer to series of such pairwise contractions and SVDs as a sweep.
Following a simulated measurement of $m$, which might partially collapse the
    entire state's entanglement, sweeps outward from $m$ may be used to
    propagate this collapse and reduce our space usage.

Finally, applying a single-qudit gate $U$ as a unitary transformation to qudit
    $m$ in an MPS may be performed as a local operation:
    \begin{align*}
        \Gamma_{\alpha_{m - 1} \alpha_m}^{\qty[m] i_m} \gets U_{i_m i'_m}
            \Gamma_{\alpha_{m - 1} \alpha_m}^{\qty[m] i'_m}.
    \end{align*}
This can be generalised to multiple-qudit gates acting on consecutive qudits in
    an MPS by contracting them, applying the gate, and then decomposing.
\textsc{Swap} gates may be used to rearrange the order of qudits in an MPS.

\section{Entanglement in Shor's Algorithm} \label{sec:shors_entanglement}
Since the space usage of an MPS scales with the amount of entanglement in the
    state, we should examine the entangling properties of Shor's algorithm in
    order to determine an efficient embedding of the circuit's $3l$ qubits into
    an MPS with its inherently linear connectivity.
By rewriting Eq.~\eqref{eq:upper_register} for the state of the upper register
    following measurement of the lower register $R$ in the form
    \begin{align} \label{eq:upper_state_expanded}
        \ket{\psi_\text{upper}} \propto \ket{j} + \ket{j + r} + \ket{j + 2r} +
            \cdots,
    \end{align}
    we observe that the $\alpha$ least-significant bits of each state on the
    right hand side of Eq.~\eqref{eq:upper_state_expanded} are identical, where
    $\alpha$ is the number of trailing zeroes in the binary representation of
    $r$.
Therefore, immediately prior to the measurement of $R$, these $\alpha$
    least-significant qubits of the upper register only exhibit entanglement
    with $R$ and not between themselves or other qubits of the upper register.
Also at this point in the circuit, the remaining $2 l - \alpha$ most-significant
    qubits of the upper register exhibit entanglement between themselves and
    with the lower register $R$, due to the odd factor $\beta \equiv r /
    2^\alpha$ of $r$ which cannot be localised to specific qubits.

Following measurement of the lower register, all $l$ qubits in $R$ are now
    completely separable.
Entanglement between the upper register with $R$ is consequently collapsed,
    leaving the upper register's $\alpha$ least-significant qubits completely
    separable and its remaining $2 l - \alpha$ most-significant qubits only
    entangled amongst themselves.

To examine some properties of $\alpha$, we note that $r$ divides $\lambda(N)$,
    the Carmichael function at $N$, which is defined as the smallest integer
    such that $x^{\lambda(N)} \equiv 1 \bmod N$ for all $x \in \mathbf{Z}_N^*$.
By the Chinese remainder theorem and Carmichael's theorem,
    \begin{align} \label{eq:max_r}
        \lambda(N) = \lcm(p - 1, q - 1)
    \end{align}
    for our odd, squarefree semiprime $N$.
Therefore if $d_p$ is the largest integer such that $2^{d_p} | (p - 1)$ and
    similarly for $d_q$, then $\max(d_p, d_q)$ is the largest value $\alpha$ can
    take for a specific $N$ over all choices of $a$.

Also by the Chinese remainder theorem, uniformly choosing $a \neq 1$ from
    $\mathbf{Z}_N^*$ is equivalent to independently and uniformly choosing
    $a_p \in \mathbf{Z}_p^*$ and $a_q \in \mathbf{Z}_q^*$ so that
    $a_p, a_q \neq 1$.
Therefore, by \cite[\textit{Lemma A4.12}]{Nielsen:2011:QCQ:1972505},
    \begin{align} \label{eq:max_alpha}
        \Pr(\alpha = \max(d_p, d_q)) \geq \frac{1}{2}
    \end{align}
    asymptotically for randomly and uniformly chosen $a \neq 1$.

Furthermore, Dirichlet's theorem implies that there is asymptotically a
    $2^{1 - m}$ probability that $2^m | (p - 1)$ for randomly chosen prime $p$.
Therefore, $d_p$ is distributed geometrically such that $\Pr(d_p = m) = 2^{-m}$.
If $p$ and $q$ are independently chosen, $d_p$ and $d_q$ are i.i.ds whose
    maximum has an expected value
    \begin{align} \label{eq:exp_alpha}
        \mathbf{E}(\max(d_p, d_q)) = \frac{8}{3}
    \end{align}
    by \cite[Eq. (2.6)]{Szpankowski1990}.

This analysis suggests a further partitioning of the upper register.
From now, we will refer to $A$ as the set of $\alpha$ least-significant qubits
    in the upper register and $B$ as the remaining $2 l - \alpha$
    most-significant qubits of the upper register.

\section{MPS Implementation} \label{sec:mps_impl}
\begin{figure*}[htb]
    \epsfig{file=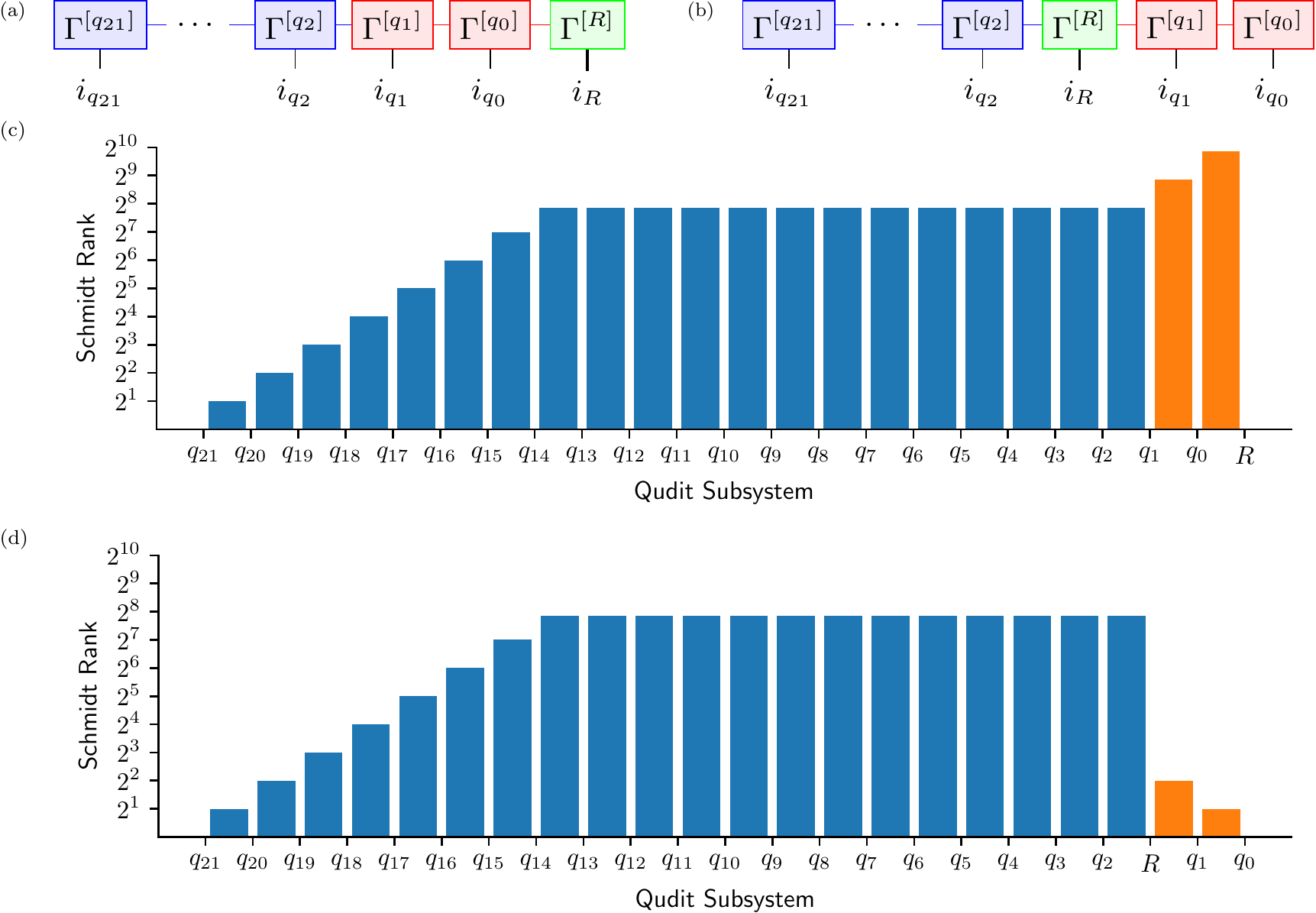,width=\textwidth,clip=}
    \caption{
        Comparison between MPS qudit layouts for example parameters $l = 11$,
            $N = 1943$, and $a = 2$, resulting in $r = 924 = 231 \times 2^2$.
        (a) and (b) show tensor network diagrams \cite{ORUS2014117} for the
            static and dynamic qudit layouts respectively, where red (blue)
            tensors correspond to tensors in $A$ ($B$) and the green tensor
            represents the lower-register qudit $R$.
        (c) and (d) show the Schmidt ranks across each bipartition in the
            static and dynamic layouts immediately following application of the
            controlled exponentiated $U$ gates, with significant reductions
            across the highlighted bipartitions.
    }
    \label{fig:shor_mps}
\end{figure*}

\subsection{Previous approaches}
High-level simulations of Shor's algorithm were performed in \cite{Wang2017}
    with a static MPS qudit ordering that places the upper-register qubits on
    one side of the lower register $R$, per the example in
    Fig.~\ref{fig:shor_mps}(a).
To briefly summarise this static method, the $l$ qubits of $R$ were kept
    contracted as a single qudit.
With $R$ effectively stored as a 1-dimensional qudit in the state $\ket{1}$,
    each of the $2 l$ upper-register qubits $q_i$ for $0 \leq i < 2 l$ is
    entangled with $R$ by applying the following procedure to each $q_i$ in
    order of descending $i$:
\begin{itemize}
    \item An initially separable qubit $q_i$ is inserted between the previous
        upper-register qubit and $R$ by altering the bond sizes appropriately
        and is set to the state $(\ket{0} + \ket{1}) / \sqrt{2}$, representing
        the starting round of Hadamard gates.
    \item The gate $U^{2^i}$, which may be formed by repeated squaring of $U$,
        Eq.~\eqref{eq:modular_unitary}, is applied to $R$ and controlled by
        $q_i$.
    This involves contracting $q_i$ with $R$ beforehand and may require $R$'s
        current effective dimensionality to be increased.
    \item This contraction is then decomposed back into qubit $q_i$ and the
        lower-register qudit $R$ by using the trivial decomposition,
        Eq.~\eqref{eq:trivial_decomposition}.
    Rank minimisation through a rank-revealing decomposition is not required at
        this stage since the apparent rank from the trivial decomposition is
        equal to the Schmidt rank.
\end{itemize}
With sufficient index book-keeping, these steps may be combined into a single
    operation per qubit $q_i$.
The effect of these controlled operations between each of the $2 l$
    upper-register qubits with the lower register $R$ is to quickly initialise
    an MPS that encodes the state $\ket{\psi_\text{modexp}}$ in
    Eq.~\eqref{eq:modular_exponentiation} with the MPS qudit layout $\qty[B] :
    \qty[A] : \qty[R]$, defined in Sec.~\ref{sec:shors_entanglement}.
As all upper-register qubits are on one side of the lower-register qudit,
    applying these controlled gates ensures that the Schmidt ranks at each
    bipartition are nondecreasing toward the lower-register qudit.
This culminates in a rank of $r$ between the least-significant qubit $q_0$ of
    the upper register with the lower register $R$ \cite{PhysRevA.69.052308},
    as demonstrated in Eq.~\eqref{eq:modular_exponentiation} and
    Fig.~\ref{fig:shor_mps}(c).
Additionally, $R$ is effectively stored as an $r$-dimensional qudit at this
    point.

This MPS is not in canonical form, since the trivial decomposition,
    Eq.~\eqref{eq:trivial_decomposition}, was used instead of the SVD,
    Eq.~\eqref{eq:svd}.
As such, nonlocal measurement of the $R$ was simulated by directly calculating
    its reduced density matrix through Eq.~\eqref{eq:density_contract}, and
    upon accordingly setting a value for $R$, entanglement is collapsed to
    reduce space usage by performing a sweep outward from $R$.
Since the lower register is now separable at this stage, it is removed from the
    MPS and the remaining upper register of $2 l$ qubits encodes the state
    $\ket{\psi_\text{upper}}$ in Eq.~\eqref{eq:upper_register}.

The quantum Fourier transform (QFT) was then performed in \cite{Wang2017} by
    sequentially contracting qubits together, eventually resulting in a
    $2^{2 l}$-dimensional state vector of the upper register.
Using this full contraction ultimately limits the size of an instance of Shor's
    algorithm that may be simulated.
However, it does scale well timewise with parallelism since decompositions are
    not required.
Though in that case, for these high-level simulations, the use of a distributed
    discrete Fourier transform library \cite{Frigo:2004:FFT:989393.989457} as a
    single operation on the fully contracted state vector should result in
    greater performance.

A separate approach to simulating Shor's algorithm up to the lower-register
    measurement, making use of a tree tensor network, is detailed in
    \cite{PhysRevA.96.062322}.
Due to the strict linear layout of an MPS, this tree tensor network more evenly
    represents the multipartite entanglement induced by the controlled
    exponentiated $U$ gates at the cost of a more complicated series of tensor
    operations when performing basic tasks such as measurement or applying
    nearest-neighbour gates.
While \cite{PhysRevA.96.062322} elucidates how such a tree tensor network might
    be converted to an MPS with a more even entanglement distribution than in
    \cite{Wang2017}, we detail how such an MPS might be generated solely
    through MPS operations.

\subsection{Simulation optimisations}
Instead of returning a state vector following a simulation of Shor's algorithm,
    which ultimately negates the space savings made by using the MPS formalism,
    we performed simulations of Shor's algorithm as a sampling problem where we
    only wish to sample measurements according to the underlying probability
    distribution resulting from the circuit.
This method mimics the process of obtaining results from an actual quantum
    computer, and is therefore a task that may be performed by both classical
    and quantum machines.
Furthermore, performing simulations in this way is ideal for MPS since it
    mitigates the need to contract to a state vector, and as the entanglement
    collapse following a measurement reduces memory usage.

From the description of Shor's algorithm in Sec.~\ref{sec:shors_algorithm}, the
    coefficients in all quantum states prior to the QFT are real.
Therefore, we can approximately halve our time and space requirements for this
    section by storing our MPS elements as real scalars instead of complex
    scalars of equal precision, and then later converting to complex scalars
    prior to the QFT.
These savings come at an opportune time due to the amount of entanglement
    before measurement of the lower register $R$.

The centrepiece of our additional optimisations, however, is to embed the
    lower-register qudit $R$ within our MPS.
Whilst the static approach in the previous section had qudits positioned in the
    order $\qty[B] : \qty[A] : \qty[R]$ as in Fig.~\ref{fig:shor_mps}(a), this
    has a significant space disadvantage since $A$ is not entangled to $B$, but
    the entanglement between $B$ and $R$ crosses through $A$ and multiplies the
    Schmidt ranks within $A$ by $\beta$.
Instead, we have chosen to perform our simulations by positioning our
    partitions as $\qty[B] : \qty[R] : \qty[A]$ in our MPS as in
    Fig.~\ref{fig:shor_mps}(b), which allows direct connectivity to $R$ from
    $A$ and $B$ simultaneously.
With this dynamic layout, the size of the MPS matrices for each of the $\alpha$
    qubits in $A$ is reduced by a factor of $\beta^2$ compared to the static
    approach as seen by comparing Figs.~\ref{fig:shor_mps}(c) and
    \ref{fig:shor_mps}(d), resulting from a factor of $\beta$ for the left and
    right bond indices of a matrix.
Therefore, the dynamic qudit layout refines classification of the difficulty in
    simulating Shor's algorithm to include the factors of $r$, rather than just
    $r$ per the static layout.

Since simulations should not depend on prior knowledge of the values $r$,
    $\alpha$ or $\beta$, our layout requires us to dynamically detect the
    transition from qubits in $B$ to qubits in $A$ when applying the controlled
    exponentiated $U$ gates.
In Sec.~\ref{sec:shors_entanglement}, we noted that $B$ was entangled with $R$
    prior to the lower-register measurement, and that the qubits within $B$
    were entangled amongst one another.
Therefore, as the controlled exponentiated $U$ gates are sequentially operated
    on qubits in order of descending significance, the Schmidt ranks with $R$
    will temporarily stop increasing (at a value of $\beta$) when this amount
    of entanglement has been reached.
Only when we start operating on qubits in $A$ does the entanglement to $R$
    begin to increase again, as seen in Fig.~\ref{fig:shor_mps}(c).
By detecting this plateau and subsequent rise in the Schmidt ranks, we can
    detect the boundary between $A$ and $B$ and begin to place qubits from $A$
    on the opposite side of $R$.

At this point, we also note from Eq.~\eqref{eq:max_alpha} that the choice of $a$
    has probability at least one half to result in the maximum $\alpha$
    permitted by semiprime $N$.
Therefore, classical simulations exceeding a given memory limit may be retried a
    constant number of times with different values of $a$ before this maximum
    $\alpha$ is expected to be observed.
Furthermore, from Eq.~\eqref{eq:exp_alpha}, when the choice of $a$ results in
    the maximum $\alpha$, $A$ is expected to contain $8 / 3$ qubits across all
    choices of $N$.
In our dynamic qudit layout where the size of each MPS matrix in $A$ is reduced
    by a factor of $\beta^2$, this may result in significant space savings
    compared to the static layout.

When sufficiently many controlled exponentiated $U$ gates have been operated to
    initialise the systems $\qty[B] : \qty[R]$, we perform a right sweep as
    required for the local measurement of $R$.
As each of the $\alpha$ qubits in $A$ is interacted with $R$, their Schmidt
    ranks toward $R$ must sequentially double to reach a lower-register
    occupancy of $r = \beta \times 2^\alpha$, given the contribution of $\beta$
    from $B$.
This is demonstrated in Fig.~\ref{fig:shor_mps}(d).
Use of the trivial decomposition Eq.~\eqref{eq:trivial_decomposition} with
    careful normalisation removes the need for the left sweep across $A$.
$R$ may then be measured using Eq.~\eqref{eq:density_local}.
Sweeps are then performed outward from this site to collapse entanglement
    involving $R$ and with $R$ now separable, it is removed to leave an MPS
    consisting of systems $\qty[B] : \qty[A]$.
Since the only remaining entanglement exists between the qubits of $B$, the
    highest Schmidt rank within the MPS at this stage is $\beta$, and the
    qubits of $A$ are completely separable at this point.

We then applied the linear nearest-neighbour (LNN) circuit for the QFT to our
    remaining qubits, as shown in Figs.~\ref{fig:shor}(a) and
    \ref{fig:shor}(b).
Since each controlled phase gate is immediately followed by a \textsc{Swap} on
    the same qubits, we applied these as a combined operation.
We note that the structure of the LNN QFT circuit allows us to measure any
    qubits that no longer require further operations.
By performing these measurements as soon as possible and sweeping afterwards,
    we help mitigate any extra entanglement introduced during the progression
    of the QFT circuit.
The final result is a completely separable MPS, rather than a full state
    vector, of the upper register corresponding to a value distributed
    according to Eq. \eqref{eq:dist}.

\section{Benchmarks} \label{sec:benchmarks}
\begin{table}[htb]
    \begin{adjustbox}{width=\textwidth}
        \begin{tabular}{cccccccccc}
        \hline
        \hline
        $l$ & $r$ & $\alpha$ & $\beta$ & $n_\text{proc}$ & $t_U$ &
            $t_\text{meas}$ & $t_\text{QFT}$ & $t_\text{total}$ & Advantage
            ($\times$) \\
        \hline
        13    & 3870 & 1 & 1935
              & 4     & 118   & 24        & 288       & 420  & 4.5 \\
        & & & & 16    & 48    & 10        & 105       & 163  & 3.2 \\
        \hline
        14    & 8036 & 2 & 2009
              & 4     & 146   & 38        & 585       & 769  & 4.6 \\
        & & & & 16    & 58    & 15        & 205       & 278  & 3.3 \\
        \hline
        15    & 16104 & 3 & 2013
              & 4     & 157   & 48        & 970       & 1175 & 8.0 \\
        & & & & 16    & 63    & 19        & 339       & 421  &     \\
        \hline
        \hline
        \end{tabular}
    \end{adjustbox}
    \caption{ \label{tab:shor_spartan}
        Benchmarks for simulating Shor's algorithm with our optimisations, with
            times for the exponentiated $U$ gates, measurement, and QFT listed
            in seconds.
        With $n_\text{proc}$ cores, we simulated the three cases $l = 13$, $N =
            8189$, $a = 10$; $l = 14$, $N = 16351$, $a = 2$; and $l = 15$, $N =
            32663$, $a = 6$, and compared the total times to the static
            approach with full state vector contraction presented in
            \cite{Wang2017} to obtain the relative advantage.
    }
\end{table}

\begin{table}[htb]
    \begin{adjustbox}{width=\textwidth}
        \begin{tabular}{ccccccccc}
            \hline
            \hline
            $l$ & $r$ & $\alpha$ & $\beta$ & $n_\text{node}$ & $t_U$ &
                $t_\text{meas}$ & $t_\text{QFT}$ & $t_\text{total}$ \\
            \hline
            16    & 28140  & 2 & 7035  & 2   & 1538  & 353  & 4290  & 6181  \\
            \hline
            17    & 57516  & 2 & 14379 & 24  & 1694  & 406  & 4544  & 6644  \\
            \hline
            20    & 479568 & 4 & 29973 & 216 & 4271  & 1496 & 20236 & 26003 \\
            \hline
            \hline
        \end{tabular}
    \end{adjustbox}
    \caption{ \label{tab:shor_magnus}
        Larger instances, this time across multiple nodes of a supercomputer.
        Times for the various stages are again listed in seconds.
        Each node has 24 cores and $\SI{64}{\giga\byte}$ of RAM.
        With $n_\text{node}$ nodes, we simulated the three cases $l = 16$, $N =
            56759$, $a = 2$; $l = 17$, $N = 124631$, $a = 2$; and $l = 20$, $N
            = 961307$, $a = 5$.
    }
\end{table}

Our simulations were implemented using the Elemental
    \cite{Poulson:2013:ENF:2427023.2427030} library for distributed-memory
    linear algebra, mainly due to its divide and conquer SVD
    \cite{Jessup:1994:PAC:181594.181618}.
All results were obtained in double precision, with $\SI{64}{\bit}$ real
    scalars before the quantum Fourier transform and $\SI{128}{\bit}$ complex
    scalars during.
Square process grids were used and elements follow a two-dimensional
    element-wise cyclic distribution.

We began by simulating Shor's algorithm with the same parameters $(l, N, a)$
    benchmarked in \cite{Wang2017}.
Our results in Table~\ref{tab:shor_spartan} were obtained on a single Intel
    Xeon E5-2683 v4 with a base core clock of $\SI{2.10}{\giga\hertz}$, using
    the Intel Parallel Studio XE suite for compilers and intraprocess BLAS and
    LAPACK \cite{laug}, and Open MPI \cite{Graham2006} for our MPI
    implementation.
We also limited the space usage for these results to $\SI{8}{\giga\byte}$ of
    RAM to showcase our optimisations against the $\SI{16}{\giga\byte}$
    required in \cite{Wang2017}.
Despite using a processor of lower clock speed, our implementation appears to
    result in a significant overall performance increase.

The parameters $(l, N, a)$ chosen in \cite{Wang2017} produce values of $r$
    equal to its upper bound of $\lcm(p - 1, q - 1)$ from Eq.~\eqref{eq:max_r},
    since the static layout bases difficulty on the size of $r$.
Our dynamic approach classifies difficulty according to the factors of $r$,
    especially the odd factor $\beta$ of $r$ which is similar across the three
    cases in Table~\ref{tab:shor_spartan}.
This is somewhat reflected in our results by our increasing time advantage over
    the results of \cite{Wang2017}.

Though our individual timings for the application of the exponentiated $U$
    gates and the QFT appear slower than the results of \cite{Wang2017}, this
    is due to our $t_U$ being used to record the time to perform the sweep
    prior to lower-register measurement and us choosing to use the LNN QFT with
    interleaved measurements respectively.
Our time savings made during the lower-register measurement more than makes up
    for this.
Additionally, whilst \cite{Wang2017} is able to claim $1 / n_\text{proc}$ time
    scaling in their results due to the parallelism of MPS contraction, we
    relied more on MPS decomposition through the SVD to keep our memory
    requirements low.
It would appear then that the SVD scales differently with $n_\text{proc}$ since
    we do not observe $1 / n_\text{proc}$ time scaling.

To benchmark a truly distributed implementation, we also performed some
    simulations across multiple nodes on Magnus \cite{pawsey}, a Cray XC40
    supercomputer with 24 cores at $\SI{2.60}{\giga\hertz}$ and
    $\SI{64}{\giga\byte}$ of RAM per node.
Our results in Table~\ref{tab:shor_magnus} were run using the GNU Compiler
    Collection suite for compilers, OpenBLAS
    \cite{Wang:2013:AAG:2503210.2503219} for intranodal BLAS and LAPACK, and
    Cray's proprietary MPI implementation.

For our flagship 60-qubit $l = 20$ instance, we chose the parameters $N =
    961307$ and $a = 5$.
These parameters were specifically chosen to highlight the differences between
    the static and dynamic methods, by having the maximum period $r = 479568$
    permitted by $N$, but with $\alpha = 4$ higher than expected.
Since $r$ here is relatively high, it would be infeasible to simulate this with
    the static method even if the final MPS contraction was not performed.
In our simulation, the $\alpha = 4$ qubits in $A$ significantly decrease the
    amount of resources required, just from an understanding of the entangling
    properties of Shor's algorithm.
As such, we were able to perform a high-level simulation of sampling a
    measurement from Shor's algorithm on 60 qubits using a total of 216 nodes,
    5184 cores, and $\SI{13.824}{\tera\byte}$ of memory within $\SI{8}{\hour}$.
To the best of our knowledge, this is the largest high-level simulation of
    Shor's algorithm.

\section{Conclusion}
We performed classical simulations of Shor's algorithm as a sampling problem
    using a dynamic MPS qudit layout.
Compared to previous approaches that relied on static qudit layouts, our
    approach better maps the entanglement induced by the circuit for Shor's
    algorithm onto a system with linear connectivity as represented by an MPS.
The use of this dynamic layout also refines classification of the difficulty in
    simulating Shor's algorithm not only to include the size of the period $r$,
    but also its factors.
Furthermore, by simulating Shor's algorithm as a sampling problem, we were able
    to take advantage of measurement to collapse entanglement within the MPS.
This reduces space usage during the quantum Fourier transform with respect to
    contracting MPS matrices into a full state vector.

In particular, we found that asymptotically, on average, the power of 2 that
    divides $r$ is $8 / 3$.
This number of qubits would become completely separable following measurement of
    the lower register, greatly reducing the simulation difficulty.
We also note that instances with semiprime $N = p q$ such that $p - 1$ or
    $q - 1$ is divisible by a high power of 2 are especially easy to simulate
    if $r$ does not possess a correspondingly large odd factor.

Our optimisations resulted in significant time and space savings for instances
    with up to 45 qubits when compared to previous benchmarks using the static
    qudit layout with full state-vector contraction.
Through the use of supercomputing resources, we were able to simulate a
    60-qubit instance of Shor's algorithm with high $r$, rendering it
    infeasible via the static approach.
In terms of the number of qubits, this represents one of the largest
    simulations of a nontrivial quantum circuit ever performed.

\begin{acknowledgements}
This work was supported by the Australian Research Council (ARC) under the
    Centre of Excellence scheme (Project No. CE110001027).
Computational resources were provided by the Pawsey Supercomputing Centre with
    funding from the Australian Government and the Government of Western
    Australia.
\end{acknowledgements}

\bibliographystyle{apsrev4-1}
\bibliography{main}

\begin{thebibliography}{32}%
\makeatletter
\providecommand \@ifxundefined [1]{%
 \@ifx{#1\undefined}
}%
\providecommand \@ifnum [1]{%
 \ifnum #1\expandafter \@firstoftwo
 \else \expandafter \@secondoftwo
 \fi
}%
\providecommand \@ifx [1]{%
 \ifx #1\expandafter \@firstoftwo
 \else \expandafter \@secondoftwo
 \fi
}%
\providecommand \natexlab [1]{#1}%
\providecommand \enquote  [1]{``#1''}%
\providecommand \bibnamefont  [1]{#1}%
\providecommand \bibfnamefont [1]{#1}%
\providecommand \citenamefont [1]{#1}%
\providecommand \href@noop [0]{\@secondoftwo}%
\providecommand \href [0]{\begingroup \@sanitize@url \@href}%
\providecommand \@href[1]{\@@startlink{#1}\@@href}%
\providecommand \@@href[1]{\endgroup#1\@@endlink}%
\providecommand \@sanitize@url [0]{\catcode `\\12\catcode `\$12\catcode
  `\&12\catcode `\#12\catcode `\^12\catcode `\_12\catcode `\%12\relax}%
\providecommand \@@startlink[1]{}%
\providecommand \@@endlink[0]{}%
\providecommand \url  [0]{\begingroup\@sanitize@url \@url }%
\providecommand \@url [1]{\endgroup\@href {#1}{\urlprefix }}%
\providecommand \urlprefix  [0]{URL }%
\providecommand \Eprint [0]{\href }%
\providecommand \doibase [0]{http://dx.doi.org/}%
\providecommand \selectlanguage [0]{\@gobble}%
\providecommand \bibinfo  [0]{\@secondoftwo}%
\providecommand \bibfield  [0]{\@secondoftwo}%
\providecommand \translation [1]{[#1]}%
\providecommand \BibitemOpen [0]{}%
\providecommand \bibitemStop [0]{}%
\providecommand \bibitemNoStop [0]{.\EOS\space}%
\providecommand \EOS [0]{\spacefactor3000\relax}%
\providecommand \BibitemShut  [1]{\csname bibitem#1\endcsname}%
\let\auto@bib@innerbib\@empty
\bibitem [{\citenamefont {Aaronson}\ and\ \citenamefont
  {Arkhipov}(2011)}]{Aaronson:2011:CCL:1993636.1993682}%
  \BibitemOpen
  \bibfield  {author} {\bibinfo {author} {\bibfnamefont {S.}~\bibnamefont
  {Aaronson}}\ and\ \bibinfo {author} {\bibfnamefont {A.}~\bibnamefont
  {Arkhipov}},\ }in\ \href {\doibase 10.1145/1993636.1993682} {\emph {\bibinfo
  {booktitle} {STOC '11 Proceedings of the Forty-third Annual ACM Symposium on
  Theory of Computing}}}\ (\bibinfo  {publisher} {ACM},\ \bibinfo {address}
  {New York},\ \bibinfo {year} {2011})\ pp.\ \bibinfo {pages}
  {333--342}\BibitemShut {NoStop}%
\bibitem [{\citenamefont {Bremner}\ \emph {et~al.}(2011)\citenamefont
  {Bremner}, \citenamefont {Jozsa},\ and\ \citenamefont {Shepherd}}]{IQP}%
  \BibitemOpen
  \bibfield  {author} {\bibinfo {author} {\bibfnamefont {M.~J.}\ \bibnamefont
  {Bremner}}, \bibinfo {author} {\bibfnamefont {R.}~\bibnamefont {Jozsa}}, \
  and\ \bibinfo {author} {\bibfnamefont {D.~J.}\ \bibnamefont {Shepherd}},\
  }\href {\doibase 10.1098/rspa.2010.0301} {\bibfield  {journal} {\bibinfo
  {journal} {Proc. Royal Soc. A}\ }\textbf {\bibinfo {volume} {467}},\ \bibinfo
  {pages} {459} (\bibinfo {year} {2011})}\BibitemShut {NoStop}%
\bibitem [{\citenamefont {Boixo}\ \emph {et~al.}(2018)\citenamefont {Boixo},
  \citenamefont {Isakov}, \citenamefont {Smelyanskiy}, \citenamefont {Babbush},
  \citenamefont {Ding}, \citenamefont {Jiang}, \citenamefont {Bremner},
  \citenamefont {Martinis},\ and\ \citenamefont {Neven}}]{2016arXiv160800263B}%
  \BibitemOpen
  \bibfield  {author} {\bibinfo {author} {\bibfnamefont {S.}~\bibnamefont
  {Boixo}}, \bibinfo {author} {\bibfnamefont {S.~V.}\ \bibnamefont {Isakov}},
  \bibinfo {author} {\bibfnamefont {V.~N.}\ \bibnamefont {Smelyanskiy}},
  \bibinfo {author} {\bibfnamefont {R.}~\bibnamefont {Babbush}}, \bibinfo
  {author} {\bibfnamefont {N.}~\bibnamefont {Ding}}, \bibinfo {author}
  {\bibfnamefont {Z.}~\bibnamefont {Jiang}}, \bibinfo {author} {\bibfnamefont
  {M.~J.}\ \bibnamefont {Bremner}}, \bibinfo {author} {\bibfnamefont {J.~M.}\
  \bibnamefont {Martinis}}, \ and\ \bibinfo {author} {\bibfnamefont
  {H.}~\bibnamefont {Neven}},\ }\href {\doibase 10.1038/s41567-018-0124-x}
  {\bibfield  {journal} {\bibinfo  {journal} {Nat. Phys.}\ }\textbf {\bibinfo
  {volume} {14}},\ \bibinfo {pages} {595} (\bibinfo {year} {2018})}\BibitemShut
  {NoStop}%
\bibitem [{\citenamefont {Bravyi}\ and\ \citenamefont
  {Gosset}(2016)}]{PhysRevLett.116.250501}%
  \BibitemOpen
  \bibfield  {author} {\bibinfo {author} {\bibfnamefont {S.}~\bibnamefont
  {Bravyi}}\ and\ \bibinfo {author} {\bibfnamefont {D.}~\bibnamefont
  {Gosset}},\ }\href {\doibase 10.1103/PhysRevLett.116.250501} {\bibfield
  {journal} {\bibinfo  {journal} {Phys. Rev. Lett.}\ }\textbf {\bibinfo
  {volume} {116}},\ \bibinfo {pages} {250501} (\bibinfo {year}
  {2016})}\BibitemShut {NoStop}%
\bibitem [{\citenamefont {Pednault}\ \emph {et~al.}(2017)\citenamefont
  {Pednault}, \citenamefont {Gunnels}, \citenamefont {Nannicini}, \citenamefont
  {Horesh}, \citenamefont {Magerlein}, \citenamefont {Solomonik},\ and\
  \citenamefont {Wisnieff}}]{2017arXiv171005867P}%
  \BibitemOpen
  \bibfield  {author} {\bibinfo {author} {\bibfnamefont {E.}~\bibnamefont
  {Pednault}}, \bibinfo {author} {\bibfnamefont {J.~A.}\ \bibnamefont
  {Gunnels}}, \bibinfo {author} {\bibfnamefont {G.}~\bibnamefont {Nannicini}},
  \bibinfo {author} {\bibfnamefont {L.}~\bibnamefont {Horesh}}, \bibinfo
  {author} {\bibfnamefont {T.}~\bibnamefont {Magerlein}}, \bibinfo {author}
  {\bibfnamefont {E.}~\bibnamefont {Solomonik}}, \ and\ \bibinfo {author}
  {\bibfnamefont {R.}~\bibnamefont {Wisnieff}},\ }\href@noop {} {\  (\bibinfo
  {year} {2017})},\ \Eprint {http://arxiv.org/abs/1710.05867}
  {arXiv:1710.05867} \BibitemShut {NoStop}%
\bibitem [{\citenamefont {Shor}(1997)}]{doi:10.1137/S0097539795293172}%
  \BibitemOpen
  \bibfield  {author} {\bibinfo {author} {\bibfnamefont {P.~W.}\ \bibnamefont
  {Shor}},\ }\href {\doibase 10.1137/S0097539795293172} {\bibfield  {journal}
  {\bibinfo  {journal} {SIAM J. Comput.}\ }\textbf {\bibinfo {volume} {26}},\
  \bibinfo {pages} {1484} (\bibinfo {year} {1997})}\BibitemShut {NoStop}%
\bibitem [{\citenamefont {Rivest}\ \emph {et~al.}(1978)\citenamefont {Rivest},
  \citenamefont {Shamir},\ and\ \citenamefont
  {Adleman}}]{Rivest:1978:MOD:359340.359342}%
  \BibitemOpen
  \bibfield  {author} {\bibinfo {author} {\bibfnamefont {R.~L.}\ \bibnamefont
  {Rivest}}, \bibinfo {author} {\bibfnamefont {A.}~\bibnamefont {Shamir}}, \
  and\ \bibinfo {author} {\bibfnamefont {L.}~\bibnamefont {Adleman}},\ }\href
  {\doibase 10.1145/359340.359342} {\bibfield  {journal} {\bibinfo  {journal}
  {Commun. ACM}\ }\textbf {\bibinfo {volume} {21}},\ \bibinfo {pages} {120}
  (\bibinfo {year} {1978})}\BibitemShut {NoStop}%
\bibitem [{\citenamefont {Dierks}\ and\ \citenamefont
  {Rescorla}(2008)}]{RFC5246}%
  \BibitemOpen
  \bibfield  {author} {\bibinfo {author} {\bibfnamefont {T.}~\bibnamefont
  {Dierks}}\ and\ \bibinfo {author} {\bibfnamefont {E.}~\bibnamefont
  {Rescorla}},\ }\href {\doibase 10.17487/RFC5246} {\emph {\bibinfo {title}
  {{The Transport Layer Security (TLS) Protocol Version 1.2}}}},\ \bibinfo
  {type} {RFC}\ \bibinfo {number} {5246}\ (\bibinfo  {institution} {RFC
  Editor},\ \bibinfo {year} {2008})\BibitemShut {NoStop}%
\bibitem [{\citenamefont {Bernstein}(2009)}]{Bernstein2009}%
  \BibitemOpen
  \bibfield  {author} {\bibinfo {author} {\bibfnamefont {D.~J.}\ \bibnamefont
  {Bernstein}},\ }\enquote {\bibinfo {title} {{Introduction to post-quantum
  cryptography}},}\ in\ \href {\doibase 10.1007/978-3-540-88702-7_1} {\emph
  {\bibinfo {booktitle} {{Post-Quantum Cryptography}}}},\ \bibinfo {editor}
  {edited by\ \bibinfo {editor} {\bibfnamefont {D.~J.}\ \bibnamefont
  {Bernstein}}, \bibinfo {editor} {\bibfnamefont {J.}~\bibnamefont {Buchmann}},
  \ and\ \bibinfo {editor} {\bibfnamefont {E.}~\bibnamefont {Dahmen}}}\
  (\bibinfo  {publisher} {Springer Berlin Heidelberg},\ \bibinfo {address}
  {Berlin, Heidelberg},\ \bibinfo {year} {2009})\ pp.\ \bibinfo {pages}
  {1--14}\BibitemShut {NoStop}%
\bibitem [{\citenamefont {Vandersypen}\ \emph {et~al.}(2001)\citenamefont
  {Vandersypen}, \citenamefont {Steffen}, \citenamefont {Breyta}, \citenamefont
  {Yannoni}, \citenamefont {Sherwood},\ and\ \citenamefont
  {Chuang}}]{Vandersypen2001}%
  \BibitemOpen
  \bibfield  {author} {\bibinfo {author} {\bibfnamefont {L.~M.~K.}\
  \bibnamefont {Vandersypen}}, \bibinfo {author} {\bibfnamefont
  {M.}~\bibnamefont {Steffen}}, \bibinfo {author} {\bibfnamefont
  {G.}~\bibnamefont {Breyta}}, \bibinfo {author} {\bibfnamefont {C.~S.}\
  \bibnamefont {Yannoni}}, \bibinfo {author} {\bibfnamefont {M.~H.}\
  \bibnamefont {Sherwood}}, \ and\ \bibinfo {author} {\bibfnamefont {I.~L.}\
  \bibnamefont {Chuang}},\ }\href {\doibase 10.1038/414883a} {\bibfield
  {journal} {\bibinfo  {journal} {Nature}\ }\textbf {\bibinfo {volume} {414}},\
  \bibinfo {pages} {883} (\bibinfo {year} {2001})}\BibitemShut {NoStop}%
\bibitem [{\citenamefont {Lanyon}\ \emph {et~al.}(2007)\citenamefont {Lanyon},
  \citenamefont {Weinhold}, \citenamefont {Langford}, \citenamefont {Barbieri},
  \citenamefont {James}, \citenamefont {Gilchrist},\ and\ \citenamefont
  {White}}]{PhysRevLett.99.250505}%
  \BibitemOpen
  \bibfield  {author} {\bibinfo {author} {\bibfnamefont {B.~P.}\ \bibnamefont
  {Lanyon}}, \bibinfo {author} {\bibfnamefont {T.~J.}\ \bibnamefont
  {Weinhold}}, \bibinfo {author} {\bibfnamefont {N.~K.}\ \bibnamefont
  {Langford}}, \bibinfo {author} {\bibfnamefont {M.}~\bibnamefont {Barbieri}},
  \bibinfo {author} {\bibfnamefont {D.~F.~V.}\ \bibnamefont {James}}, \bibinfo
  {author} {\bibfnamefont {A.}~\bibnamefont {Gilchrist}}, \ and\ \bibinfo
  {author} {\bibfnamefont {A.~G.}\ \bibnamefont {White}},\ }\href {\doibase
  10.1103/PhysRevLett.99.250505} {\bibfield  {journal} {\bibinfo  {journal}
  {Phys. Rev. Lett.}\ }\textbf {\bibinfo {volume} {99}},\ \bibinfo {pages}
  {250505} (\bibinfo {year} {2007})}\BibitemShut {NoStop}%
\bibitem [{\citenamefont {Lu}\ \emph {et~al.}(2007)\citenamefont {Lu},
  \citenamefont {Browne}, \citenamefont {Yang},\ and\ \citenamefont
  {Pan}}]{PhysRevLett.99.250504}%
  \BibitemOpen
  \bibfield  {author} {\bibinfo {author} {\bibfnamefont {C.-Y.}\ \bibnamefont
  {Lu}}, \bibinfo {author} {\bibfnamefont {D.~E.}\ \bibnamefont {Browne}},
  \bibinfo {author} {\bibfnamefont {T.}~\bibnamefont {Yang}}, \ and\ \bibinfo
  {author} {\bibfnamefont {J.-W.}\ \bibnamefont {Pan}},\ }\href {\doibase
  10.1103/PhysRevLett.99.250504} {\bibfield  {journal} {\bibinfo  {journal}
  {Phys. Rev. Lett.}\ }\textbf {\bibinfo {volume} {99}},\ \bibinfo {pages}
  {250504} (\bibinfo {year} {2007})}\BibitemShut {NoStop}%
\bibitem [{\citenamefont {Lucero}\ \emph {et~al.}(2012)\citenamefont {Lucero},
  \citenamefont {Barends}, \citenamefont {Chen}, \citenamefont {Kelly},
  \citenamefont {Mariantoni}, \citenamefont {Megrant}, \citenamefont
  {O'Malley}, \citenamefont {Sank}, \citenamefont {Vainsencher}, \citenamefont
  {Wenner}, \citenamefont {White}, \citenamefont {Yin}, \citenamefont
  {Cleland},\ and\ \citenamefont {Martinis}}]{Lucero2012}%
  \BibitemOpen
  \bibfield  {author} {\bibinfo {author} {\bibfnamefont {E.}~\bibnamefont
  {Lucero}}, \bibinfo {author} {\bibfnamefont {R.}~\bibnamefont {Barends}},
  \bibinfo {author} {\bibfnamefont {Y.}~\bibnamefont {Chen}}, \bibinfo {author}
  {\bibfnamefont {J.}~\bibnamefont {Kelly}}, \bibinfo {author} {\bibfnamefont
  {M.}~\bibnamefont {Mariantoni}}, \bibinfo {author} {\bibfnamefont
  {A.}~\bibnamefont {Megrant}}, \bibinfo {author} {\bibfnamefont
  {P.}~\bibnamefont {O'Malley}}, \bibinfo {author} {\bibfnamefont
  {D.}~\bibnamefont {Sank}}, \bibinfo {author} {\bibfnamefont {A.}~\bibnamefont
  {Vainsencher}}, \bibinfo {author} {\bibfnamefont {J.}~\bibnamefont {Wenner}},
  \bibinfo {author} {\bibfnamefont {T.}~\bibnamefont {White}}, \bibinfo
  {author} {\bibfnamefont {Y.}~\bibnamefont {Yin}}, \bibinfo {author}
  {\bibfnamefont {A.~N.}\ \bibnamefont {Cleland}}, \ and\ \bibinfo {author}
  {\bibfnamefont {J.~M.}\ \bibnamefont {Martinis}},\ }\href {\doibase
  10.1038/nphys2385} {\bibfield  {journal} {\bibinfo  {journal} {Nat. Phys.}\
  }\textbf {\bibinfo {volume} {8}},\ \bibinfo {pages} {719} (\bibinfo {year}
  {2012})}\BibitemShut {NoStop}%
\bibitem [{\citenamefont {Wang}\ \emph {et~al.}(2017)\citenamefont {Wang},
  \citenamefont {Hill},\ and\ \citenamefont {Hollenberg}}]{Wang2017}%
  \BibitemOpen
  \bibfield  {author} {\bibinfo {author} {\bibfnamefont {D.~S.}\ \bibnamefont
  {Wang}}, \bibinfo {author} {\bibfnamefont {C.~D.}\ \bibnamefont {Hill}}, \
  and\ \bibinfo {author} {\bibfnamefont {L.~C.~L.}\ \bibnamefont
  {Hollenberg}},\ }\href {\doibase 10.1007/s11128-017-1587-x} {\bibfield
  {journal} {\bibinfo  {journal} {Quantum Inf. Process.}\ }\textbf {\bibinfo
  {volume} {16}},\ \bibinfo {pages} {176} (\bibinfo {year} {2017})}\BibitemShut
  {NoStop}%
\bibitem [{\citenamefont {Vidal}(2003)}]{PhysRevLett.91.147902}%
  \BibitemOpen
  \bibfield  {author} {\bibinfo {author} {\bibfnamefont {G.}~\bibnamefont
  {Vidal}},\ }\href {\doibase 10.1103/PhysRevLett.91.147902} {\bibfield
  {journal} {\bibinfo  {journal} {Phys. Rev. Lett.}\ }\textbf {\bibinfo
  {volume} {91}},\ \bibinfo {pages} {147902} (\bibinfo {year}
  {2003})}\BibitemShut {NoStop}%
\bibitem [{\citenamefont {Or\'us}(2014)}]{ORUS2014117}%
  \BibitemOpen
  \bibfield  {author} {\bibinfo {author} {\bibfnamefont {R.}~\bibnamefont
  {Or\'us}},\ }\href {\doibase 10.1016/j.aop.2014.06.013} {\bibfield  {journal}
  {\bibinfo  {journal} {Ann. Phys. (N. Y.)}\ }\textbf {\bibinfo {volume}
  {349}},\ \bibinfo {pages} {117} (\bibinfo {year} {2014})}\BibitemShut
  {NoStop}%
\bibitem [{\citenamefont {Schollw\"ock}(2011)}]{SCHOLLWOCK201196}%
  \BibitemOpen
  \bibfield  {author} {\bibinfo {author} {\bibfnamefont {U.}~\bibnamefont
  {Schollw\"ock}},\ }\href {\doibase 10.1016/j.aop.2010.09.012} {\bibfield
  {journal} {\bibinfo  {journal} {Ann. Phys. (N. Y.)}\ }\textbf {\bibinfo
  {volume} {326}},\ \bibinfo {pages} {96} (\bibinfo {year} {2011})}\BibitemShut
  {NoStop}%
\bibitem [{\citenamefont {Vidal}(2007{\natexlab{a}})}]{PhysRevLett.98.070201}%
  \BibitemOpen
  \bibfield  {author} {\bibinfo {author} {\bibfnamefont {G.}~\bibnamefont
  {Vidal}},\ }\href {\doibase 10.1103/PhysRevLett.98.070201} {\bibfield
  {journal} {\bibinfo  {journal} {Phys. Rev. Lett.}\ }\textbf {\bibinfo
  {volume} {98}},\ \bibinfo {pages} {070201} (\bibinfo {year}
  {2007}{\natexlab{a}})}\BibitemShut {NoStop}%
\bibitem [{\citenamefont {Woolfe}\ \emph {et~al.}(2017)\citenamefont {Woolfe},
  \citenamefont {Hill},\ and\ \citenamefont
  {Hollenberg}}]{2014arXiv1406.0931W}%
  \BibitemOpen
  \bibfield  {author} {\bibinfo {author} {\bibfnamefont {K.~J.}\ \bibnamefont
  {Woolfe}}, \bibinfo {author} {\bibfnamefont {C.~D.}\ \bibnamefont {Hill}}, \
  and\ \bibinfo {author} {\bibfnamefont {L.~C.~L.}\ \bibnamefont
  {Hollenberg}},\ }\href {\doibase 10.26421/QIC17.1-2} {\bibfield  {journal}
  {\bibinfo  {journal} {Quantum Inf. Comput.}\ }\textbf {\bibinfo {volume}
  {17}},\ \bibinfo {pages} {1} (\bibinfo {year} {2017})}\BibitemShut {NoStop}%
\bibitem [{\citenamefont {Verstraete}\ and\ \citenamefont
  {Cirac}(2004)}]{peps}%
  \BibitemOpen
  \bibfield  {author} {\bibinfo {author} {\bibfnamefont {F.}~\bibnamefont
  {Verstraete}}\ and\ \bibinfo {author} {\bibfnamefont {J.~I.}\ \bibnamefont
  {Cirac}},\ }\href@noop {} {\  (\bibinfo {year} {2004})},\ \Eprint
  {http://arxiv.org/abs/cond-mat/0407066} {arXiv:cond-mat/0407066} \BibitemShut
  {NoStop}%
\bibitem [{\citenamefont {Vidal}(2007{\natexlab{b}})}]{PhysRevLett.99.220405}%
  \BibitemOpen
  \bibfield  {author} {\bibinfo {author} {\bibfnamefont {G.}~\bibnamefont
  {Vidal}},\ }\href {\doibase 10.1103/PhysRevLett.99.220405} {\bibfield
  {journal} {\bibinfo  {journal} {Phys. Rev. Lett.}\ }\textbf {\bibinfo
  {volume} {99}},\ \bibinfo {pages} {220405} (\bibinfo {year}
  {2007}{\natexlab{b}})}\BibitemShut {NoStop}%
\bibitem [{\citenamefont {Dumitrescu}(2017)}]{PhysRevA.96.062322}%
  \BibitemOpen
  \bibfield  {author} {\bibinfo {author} {\bibfnamefont {E.}~\bibnamefont
  {Dumitrescu}},\ }\href {\doibase 10.1103/PhysRevA.96.062322} {\bibfield
  {journal} {\bibinfo  {journal} {Phys. Rev. A}\ }\textbf {\bibinfo {volume}
  {96}},\ \bibinfo {pages} {062322} (\bibinfo {year} {2017})}\BibitemShut
  {NoStop}%
\bibitem [{\citenamefont {Nielsen}\ and\ \citenamefont
  {Chuang}(2010)}]{Nielsen:2011:QCQ:1972505}%
  \BibitemOpen
  \bibfield  {author} {\bibinfo {author} {\bibfnamefont {M.~A.}\ \bibnamefont
  {Nielsen}}\ and\ \bibinfo {author} {\bibfnamefont {I.~L.}\ \bibnamefont
  {Chuang}},\ }\href {\doibase 10.1017/CBO9780511976667} {\emph {\bibinfo
  {title} {{Quantum Computation and Quantum Information}}}},\ \bibinfo
  {edition} {{10th Anniversary}}\ ed.\ (\bibinfo  {publisher} {Cambridge
  University Press},\ \bibinfo {address} {Cambridge},\ \bibinfo {year}
  {2010})\BibitemShut {NoStop}%
\bibitem [{\citenamefont {Szpankowski}\ and\ \citenamefont
  {Rego}(1990)}]{Szpankowski1990}%
  \BibitemOpen
  \bibfield  {author} {\bibinfo {author} {\bibfnamefont {W.}~\bibnamefont
  {Szpankowski}}\ and\ \bibinfo {author} {\bibfnamefont {V.}~\bibnamefont
  {Rego}},\ }\href {\doibase 10.1007/BF02241658} {\bibfield  {journal}
  {\bibinfo  {journal} {Computing}\ }\textbf {\bibinfo {volume} {43}},\
  \bibinfo {pages} {401} (\bibinfo {year} {1990})}\BibitemShut {NoStop}%
\bibitem [{\citenamefont {Or\'us}\ and\ \citenamefont
  {Latorre}(2004)}]{PhysRevA.69.052308}%
  \BibitemOpen
  \bibfield  {author} {\bibinfo {author} {\bibfnamefont {R.}~\bibnamefont
  {Or\'us}}\ and\ \bibinfo {author} {\bibfnamefont {J.~I.}\ \bibnamefont
  {Latorre}},\ }\href {\doibase 10.1103/PhysRevA.69.052308} {\bibfield
  {journal} {\bibinfo  {journal} {Phys. Rev. A}\ }\textbf {\bibinfo {volume}
  {69}},\ \bibinfo {pages} {052308} (\bibinfo {year} {2004})}\BibitemShut
  {NoStop}%
\bibitem [{\citenamefont {Frigo}(2004)}]{Frigo:2004:FFT:989393.989457}%
  \BibitemOpen
  \bibfield  {author} {\bibinfo {author} {\bibfnamefont {M.}~\bibnamefont
  {Frigo}},\ }\href {\doibase 10.1145/989393.989457} {\bibfield  {journal}
  {\bibinfo  {journal} {SIGPLAN Not.}\ }\textbf {\bibinfo {volume} {39}},\
  \bibinfo {pages} {642} (\bibinfo {year} {2004})}\BibitemShut {NoStop}%
\bibitem [{\citenamefont {Poulson}\ \emph {et~al.}(2013)\citenamefont
  {Poulson}, \citenamefont {Marker}, \citenamefont {van~de Geijn},
  \citenamefont {Hammond},\ and\ \citenamefont
  {Romero}}]{Poulson:2013:ENF:2427023.2427030}%
  \BibitemOpen
  \bibfield  {author} {\bibinfo {author} {\bibfnamefont {J.}~\bibnamefont
  {Poulson}}, \bibinfo {author} {\bibfnamefont {B.}~\bibnamefont {Marker}},
  \bibinfo {author} {\bibfnamefont {R.~A.}\ \bibnamefont {van~de Geijn}},
  \bibinfo {author} {\bibfnamefont {J.~R.}\ \bibnamefont {Hammond}}, \ and\
  \bibinfo {author} {\bibfnamefont {N.~A.}\ \bibnamefont {Romero}},\ }\href
  {\doibase 10.1145/2427023.2427030} {\bibfield  {journal} {\bibinfo  {journal}
  {ACM Trans. Math. Softw.}\ }\textbf {\bibinfo {volume} {39}},\ \bibinfo
  {pages} {13:1} (\bibinfo {year} {2013})}\BibitemShut {NoStop}%
\bibitem [{\citenamefont {Jessup}\ and\ \citenamefont
  {Sorensen}(1994)}]{Jessup:1994:PAC:181594.181618}%
  \BibitemOpen
  \bibfield  {author} {\bibinfo {author} {\bibfnamefont {E.~R.}\ \bibnamefont
  {Jessup}}\ and\ \bibinfo {author} {\bibfnamefont {D.~C.}\ \bibnamefont
  {Sorensen}},\ }\href {\doibase 10.1137/S089547989120195X} {\bibfield
  {journal} {\bibinfo  {journal} {SIAM J. Matrix Anal. Appl.}\ }\textbf
  {\bibinfo {volume} {15}},\ \bibinfo {pages} {530} (\bibinfo {year}
  {1994})}\BibitemShut {NoStop}%
\bibitem [{\citenamefont {Anderson}\ \emph {et~al.}(1999)\citenamefont
  {Anderson}, \citenamefont {Bai}, \citenamefont {Bischof}, \citenamefont
  {Blackford}, \citenamefont {Demmel}, \citenamefont {Dongarra}, \citenamefont
  {Du~Croz}, \citenamefont {Greenbaum}, \citenamefont {Hammarling},
  \citenamefont {McKenney},\ and\ \citenamefont {Sorensen}}]{laug}%
  \BibitemOpen
  \bibfield  {author} {\bibinfo {author} {\bibfnamefont {E.}~\bibnamefont
  {Anderson}}, \bibinfo {author} {\bibfnamefont {Z.}~\bibnamefont {Bai}},
  \bibinfo {author} {\bibfnamefont {C.}~\bibnamefont {Bischof}}, \bibinfo
  {author} {\bibfnamefont {L.}~\bibnamefont {Blackford}}, \bibinfo {author}
  {\bibfnamefont {J.}~\bibnamefont {Demmel}}, \bibinfo {author} {\bibfnamefont
  {J.}~\bibnamefont {Dongarra}}, \bibinfo {author} {\bibfnamefont
  {J.}~\bibnamefont {Du~Croz}}, \bibinfo {author} {\bibfnamefont
  {A.}~\bibnamefont {Greenbaum}}, \bibinfo {author} {\bibfnamefont
  {S.}~\bibnamefont {Hammarling}}, \bibinfo {author} {\bibfnamefont
  {A.}~\bibnamefont {McKenney}}, \ and\ \bibinfo {author} {\bibfnamefont
  {D.}~\bibnamefont {Sorensen}},\ }\href {\doibase 10.1137/1.9780898719604}
  {\emph {\bibinfo {title} {{LAPACK Users' Guide}}}},\ \bibinfo {edition}
  {{Third}}\ ed.\ (\bibinfo  {publisher} {Society for Industrial and Applied
  Mathematics},\ \bibinfo {address} {Philadelphia},\ \bibinfo {year}
  {1999})\BibitemShut {NoStop}%
\bibitem [{\citenamefont {Graham}\ \emph {et~al.}(2006)\citenamefont {Graham},
  \citenamefont {Woodall},\ and\ \citenamefont {Squyres}}]{Graham2006}%
  \BibitemOpen
  \bibfield  {author} {\bibinfo {author} {\bibfnamefont {R.~L.}\ \bibnamefont
  {Graham}}, \bibinfo {author} {\bibfnamefont {T.~S.}\ \bibnamefont {Woodall}},
  \ and\ \bibinfo {author} {\bibfnamefont {J.~M.}\ \bibnamefont {Squyres}},\
  }in\ \href {\doibase 10.1007/11752578_29} {\emph {\bibinfo {booktitle}
  {{Parallel Processing and Applied Mathematics: 6th International Conference,
  PPAM 2005, Pozna\'n, Poland, September 11-14, 2005, Revised Selected
  Papers}}}},\ \bibinfo {editor} {edited by\ \bibinfo {editor} {\bibfnamefont
  {R.}~\bibnamefont {Wyrzykowski}}, \bibinfo {editor} {\bibfnamefont
  {J.}~\bibnamefont {Dongarra}}, \bibinfo {editor} {\bibfnamefont
  {N.}~\bibnamefont {Meyer}}, \ and\ \bibinfo {editor} {\bibfnamefont
  {J.}~\bibnamefont {Wa\'sniewski}}}\ (\bibinfo  {publisher} {Springer Berlin
  Heidelberg},\ \bibinfo {address} {Berlin, Heidelberg},\ \bibinfo {year}
  {2006})\ pp.\ \bibinfo {pages} {228--239}\BibitemShut {NoStop}%
\bibitem [{paw(2017)}]{pawsey}%
  \BibitemOpen
  \href@noop {} {\enquote {\bibinfo {title} {{The Pawsey Supercomputing
  Centre}},}\ }\bibinfo {howpublished} {\url{https://www.pawsey.org.au/}}
  (\bibinfo {year} {2017})\BibitemShut {NoStop}%
\bibitem [{\citenamefont {Wang}\ \emph {et~al.}(2013)\citenamefont {Wang},
  \citenamefont {Zhang}, \citenamefont {Zhang},\ and\ \citenamefont
  {Yi}}]{Wang:2013:AAG:2503210.2503219}%
  \BibitemOpen
  \bibfield  {author} {\bibinfo {author} {\bibfnamefont {Q.}~\bibnamefont
  {Wang}}, \bibinfo {author} {\bibfnamefont {X.}~\bibnamefont {Zhang}},
  \bibinfo {author} {\bibfnamefont {Y.}~\bibnamefont {Zhang}}, \ and\ \bibinfo
  {author} {\bibfnamefont {Q.}~\bibnamefont {Yi}},\ }in\ \href {\doibase
  10.1145/2503210.2503219} {\emph {\bibinfo {booktitle} {{SC '13 Proceedings of
  the International Conference on High Performance Computing, Networking,
  Storage and Analysis}}}}\ (\bibinfo  {publisher} {ACM},\ \bibinfo {address}
  {New York},\ \bibinfo {year} {2013})\ pp.\ \bibinfo {pages}
  {25:1--25:12}\BibitemShut {NoStop}%
\end{thebibliography}%

\end{document}